\documentclass{ws-mpla}
\newcommand{\kev}{keV}

\newcommand{\fe}{Fe~K$\alpha$}
\newcommand{\etal}{et al.}
\newcommand{\mcg}{MCG--6-30-15}
\newcommand{\xmm}{\textit{XMM-Newton}}
\newcommand{\chandra}{\textit{Chandra}}
\newcommand{\ga}{\,\rlap{\raise 0.5ex\hbox{$>$}}{\lower
    1.0ex\hbox{$\sim$}}\,}
\newcommand{\la}{\,\rlap{\raise 0.5ex\hbox{$<$}}{\lower
    1.0ex\hbox{$\sim$}}\,}

\begin{document}

\markboth{D.R. Ballantyne}
{The Accretion Geometry in Radio-Loud Active Galaxies}

\catchline{}{}{}{}{}

\title{THE ACCRETION GEOMETRY IN RADIO-LOUD ACTIVE GALAXIES}

\author{\footnotesize D.R. BALLANTYNE}

\address{Department of Physics, The University of Arizona, 1118 E. 4th Street\\
Tucson, Arizona 85721, USA\\
drb@physics.arizona.edu}

\maketitle

\pub{Received (Day Month Year)}{Revised (Day Month Year)}

\begin{abstract}
We review the latest attempts to determine the accretion geometry in
radio-loud active galactic nuclei (AGN). These objects, which comprise
$\sim 10-20$\% of the AGN population, produce powerful collimated radio
jets that can extend thousands of parsecs from the center of the host
galaxy. Recent multiwavelength surveys have shown that
radio-loudness is more common in low-luminosity AGN than in higher
luminosity Seyfert galaxies or quasars. These low-luminosity AGN have
small enough accretion rates that they are most likely accreting via a geometrically thick and
radiatively inefficient accretion flow. In contrast, X-ray spectroscopic
observations of three higher luminosity broad-line radio galaxies
(3C~120, 4C+74.26 and PG~1425+267) have
found evidence for an untruncated thin disk extending very close to the
black hole. These tentative detections indicate that, for this class of
radio-loud AGN, the accretion geometry is very similar to their
radio-quiet counterparts. These observations suggest that there are
three conditions to jet formation that must be satisfied: the presence
of a rapidly spinning black hole, an accretion flow with a large $H/r$
ratio, and a favorable magnetic field geometry.

\end{abstract}

\ccode{PACS Nos.: 97.10.Gz, 98.54.Cm, 98.54.Gr}

\section{Introduction}	
\label{sect:intro}
As gas and dust accretes onto a supermassive black hole at the center
of a galaxy, a significant amount of gravitational potential energy is
liberated. In a radiatively efficient accretion flow, one-half of this
energy is radiated away to infinity,\cite{ss73} allowing these active
galactic nuclei (AGN) to be observed out to vast cosmological
distances.\cite{beck01,fan01} However, there are other avenues through which the
accretion energy can be released, such as the driving of outflows or
large scale motions of the accretion flow.\cite{bb99,qg00} These processes do
not produce strong observable signatures, so observations simply
detect a drop in the radiative efficiency of the accretion flow.

Three decades ago, radio observations of AGN uncovered another pathway
for energy release by accreting black holes: highly collimated jets of
relativistic plasma extending thousands of parsecs from the center of the
galaxy.\cite{fr74,bp84,bbr84} Only $\sim10$--$20$\% of AGN fall into this
`radio-loud' category,\cite{kel89,ive02} and, despite years of study, the exact
mechanism that produces this spectacular release of energy is still
unknown. As these jets are produced at the centers of active galaxies,
it is natural to ask if there is a difference in the accretion flow
between the radio-loud and radio-quiet populations. This short review
summarizes the attempts made over the last several years to answer
this question. We begin in the next section by collecting some of the
key observational properties and correlations of radio-loud AGN,
including the various definitions of what constitutes a radio-loud
AGN. With the explosion in large-scale and multi-wavelength surveys
over the last decade, more information than ever is available on the
properties of radio-loud AGN and their host
galaxies. Section~\ref{sect:geometry} than presents the results of
measurements of the accretion geometry of radio-loud AGN and compares
them to the non-jet producing AGN. Finally, in the last section, we
collect together all the various data and models and try to answer the
question: how do black holes make jets?

\section{Definition of a Radio-Loud AGN}
\label{sect:definition}
All AGN emit at radio wavelengths at some level (i.e., there are no
`radio-silent' sources), so the radio power must be compared to the
emission at higher frequencies to determine the `radio-loudness' of a
given object. Early radio surveys of optically selected quasars
measured the radio-loudness as the ratio between the monochromatic
luminosities at 5~GHz and 4400~\AA, $R=\nu L_{\mathrm{5\ GHz}}/\nu
L_{\mathrm{4400\ A}}$.\cite{kel89,vis92,sto92,kel94} These studies found that the distribution of
quasars was roughly bimodal, with the vast majority of objects found
to be radio-quiet (i.e., $R \la
10$). An alternative definition of $R$ is a comparison between
the radio and X-ray luminosity, $R_{X}=\nu L_{\mathrm{5\ GHz}}/\nu
L_{\mathrm{2-10\ keV}}$.\cite{tw03} The advantage of this definition is that the
hard X-ray band is less sensitive to extinction and is closely related
to the accretion power. In this case, the dividing line between the
radio-loud and radio-quiet population is $\log R_X = -4.8$.

\begin{figure}[t!]
\centerline{\psfig{file=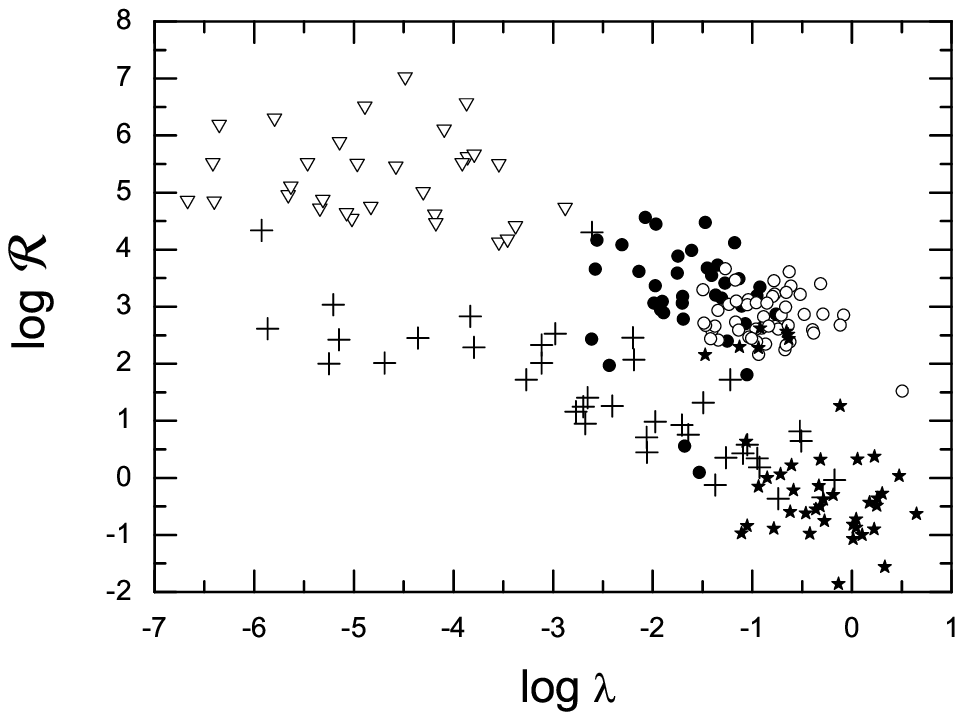,width=2.8in}
\psfig{file=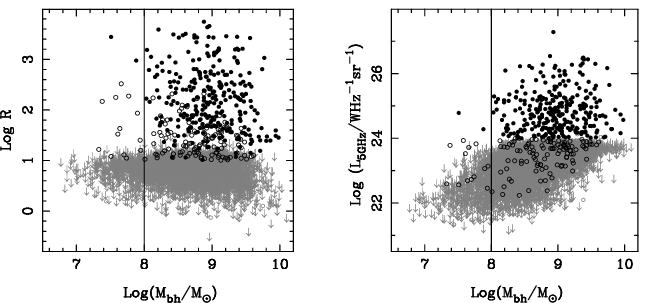, width=2.7in, trim=0 0 95 0, clip}}
\vspace*{1pt}
\caption{(Left) Plot of the radio-loudness parameter $R$ as a function
  of the Eddington ratio $\lambda \equiv
  L_{\mathrm{bol}}/L_{\mathrm{Edd}}$. The different symbols denote
  different types of AGN: filled circles $\rightarrow$ broad-line
  radio galaxies, open circles $\rightarrow$ radio-loud quasars,
  crosses $\rightarrow$ Seyfert galaxies and LINERs, open triangles
  $\rightarrow$ FR I radio galaxies and filled stars $\rightarrow$ PG quasars. Plot taken
  from Ref.~\protect\refcite{ssl07}. (Right) As in the other panel, but now plotting $R$
  against black hole mass for quasars selected from the Sloan Digital
  Sky Survey. Black points denote the radio-loud quasars, while the
  gray symbols are the radio-quiet objects. Figure taken from
  Ref.~\protect\refcite{md04}.}
\label{sikora}
\end{figure}

This relatively simple view of the so-called radio-loud/radio-quiet
dichotomy has been made significantly more complex with the advent of
deep, multi-wavelength, and wide-field imaging surveys. For example, it is no longer clear
that $R$ is bimodal, since many
radio-selected quasars fall between the original radio-loud and
radio-quiet definitions.\cite{white00,hewett01,cir03} However, other
selection techniques and surveys continue to show evidence for a
bimodality.\cite{ive02} A recent study using the Sloan
Digital Sky Survey indicates that, after years of conflicting
results,\cite{vis92,hoop95,stern00} the fraction of quasars that are
radio-loud decreases with increasing redshift and with decreasing
luminosity.\cite{jiang07} The new large surveys have also shown that if one uses the
fixed values of $R$ or $R_X$ as a dividing line, the fraction of
radio-loud AGN \emph{increases} with \emph{decreasing} Eddington
ratio.\cite{hp01,ho02,ghu06,pan07} A recent large and heterogeneous sample compiled by Sikora et
al. shows that there exists both a radio-loud and a radio-quiet
sequence at all accretion rates (see the left-hand panel of
Figure~\ref{sikora}).\cite{ssl07} These results illustrate the
limitation of using $R$ or $R_X$ in determining whether an AGN is
radio-loud or radio-quiet, since not all of these radio-loud low-luminosity AGN
produce powerful jets, the subject of this review. For example, Sgr A*
and M87 both have very similar values of $R$ showing them to be extremely
radio-loud, yet only M87 produces a powerful radio jet. The
observed increase in radio-loudness with decreasing Eddington ratio
may indicate a change in how accretion energy is directed and 
liberated around black holes. 

The last decade has also seen a massive increase in our understanding
of the host galaxies of both radio-loud and radio-quiet AGN, most
notably with the use of the radius-luminosity relation derived from
reverberation mapping to weigh the central supermassive black
hole.\cite{pw99,kas00,kas05} Various studies have concluded that while the black hole
at the heart of radio-quiet AGN can have a wide range of masses,
radio-loud AGN almost always have central black holes with masses $>
10^8$~M$_{\odot}$ (see the right-hand panel of Figure~\ref{sikora}),\cite{laor00,md02,md04,mm06} although there are dissenting views.\cite{wu02} This
result is consistent with earlier work that showed that
radio-loud AGN seem to be almost exclusively hosted by early type
galaxies (i.e., galaxies with massive bulges).\cite{xu99} Radio-quiet
AGN can reside in either spiral or elliptical
galaxies.\cite{vc91,mcl99,floyd04} Now, \textit{HST}
observations of low-luminosity AGN in early type hosts have shown that
the radio-loud sources have bulges with central cores in their
brightness profiles, while the radio-quiet ones have power-law brightness
profiles.\cite{cp06,cp07} Central cores may be formed through
dynamical effects following a merger with another massive
galaxy.\cite{milo02,rav02}

Although these observational advances are impressive, they are still
limited by the difficulty in selecting an unbiased and complete sample
of objects. This is especially difficult for AGN, where the vast majority
suffer some level of obscuration,\cite{mr95,fi99} but is doubly problematic for
radio-loud objects, as these sources are relatively scarce within the overall
population. Therefore, it is important for the above observational results to be
confirmed and clarified with a proper understanding of the
biases involved in selecting the galaxy samples.

\section{Determining the Accretion Geometry}
\label{sect:geometry}
As accretion flows in external galaxies cannot be resolved by imaging,
indirect methods such as hard X-ray spectroscopy are required to probe
the structure of the accretion disk. The variability properties of hard
X-rays from AGN shows that they are produced within the innermost regions
of the accreting material. Hard X-ray spectra of bright AGN obtained
by the \textit{Ginga} and \textit{ASCA} observatories showed that the
underlying X-ray power-law hardens at $\ga 8$~keV and often features a
Fe~K$\alpha$ fluorescent line at 6.4~keV.\cite{pou90,np94} These features are
most naturally interpreted as evidence of reprocessing (or reflection)
of the X-ray power-law in dense, relatively cold material lying out of
the line of sight.\cite{lw88,gr88,gf91} Deep observations of bright Seyfert~1
galaxies such as \mcg\ showed that the \fe\ line could be highly asymmetric
and exhibit a broad red wing.\cite{tan95,fab02,rn03} These features match those
predicted by models of X-ray illuminated accretion disks where
relativistic effects sculpt the \fe\ line into a characteristic
shape.\cite{fab89,lao91,bd04} Fitting these models to the observed X-ray spectra showed that
the line emission must be being produced well within 10~$r_g$ (where $r_g
\equiv GM/c^2$ is the gravitational radius of a black hole with mass
$M$).\cite{mill07} Thus, the reflection features observed in the hard X-ray spectra
of AGN could provide information on the extent, ionization
state,\cite{rfy99} composition,\cite{bfr02} and structure of the
accretion disk very close to the black hole.\cite{nkk00,brf01,btb04}

Unfortunately, such studies have been difficult to apply to radio-loud
AGN. First, as mentioned above, radio-loud AGN are relatively scarce
so there exists very few sources that have a high 2--10~\kev\
flux. A precise measurement of a \fe\ line profile requires a careful understanding
of the underlying continuum and thus a high signal-to-noise
spectrum. As a result, very few radio-loud AGN are bright enough in
the X-ray band to provide the high quality data necessary for detailed
spectral analysis without very long exposure times. A second
problem arises from the presence of the radio jet itself. The
relativistic plasma in the jet can produce X-rays via synchrotron, synchrotron
self-Compton or by Compton up-scattering of low energy photons from
the accretion disk. Thus, depending on the viewing angle into the central
engine and the jet dynamics, any accretion disk reflection features
may be hidden or altered by X-rays produced by the jet.\cite{page05,gmf06,sam06} 

Despite these problems, X-ray observations of a significant number of
radio-loud AGN were performed in the 1990s with \textit{ASCA},
\textit{RXTE} and \textit{BeppoSAX}.\cite{eh98,woz98,sem99,era00} The best evidence for
reflection features and \fe\ lines were found in broad-line radio
galaxies (BLRGs), radio-loud AGN whose optical spectra show similar broad
permitted lines as radio-quiet type~1 AGN. Considering the results on
BLRGs collectively it was found that they have harder X-ray
power-laws, weaker reflection strengths and smaller \fe\ equivalent
widths than their radio-quiet Seyfert~1 counterparts.\cite{gmf06} There was
no clear detection of a relativistically broadened \fe\ line from the
inner disk of a BLRG. This
result was interpreted as a change in accretion geometry between the
two classes. Specifically, the dense, geometrically thin inner
accretion disk in the radio-quiet AGN was replaced in BLRGs with a tenuous,
geometrically thick flow (such as an Advection Dominated Accretion
Flow; see Ref.~\refcite{ny95}) that would not produce strong reflection
features. Alternatively, the weak reflection features in BLRGs may also
be evidence of a highly ionized untruncated accretion
flow.\cite{bal02} A third possibility is that the reflection features
are bring diluted by the radio jet, however a recent reanalysis of \textit{BeppoSAX}
data of three BLRGs has shown that any non-thermal jet component contributes
$< 45$\% of the flux in the 2--10~\kev\ band.\cite{gp07} Of course, it
is difficult to rule out the possibility of jet dilution for every BLRG without the
presence of very high energy data.

It was hoped that the substantial increase in sensitivity afforded by
the \xmm\ and \chandra\ observatories would allow for a clearer
picture of the accretion geometry within radio-loud
AGN. Unfortunately, in many cases the results have remained frustratingly
ambiguous. The BLRGs NGC~6251, 3C 109 and 3C 111 all show evidence
of reflection and a broad \fe\ line,\cite{glio04,lewis05,min06} but the data quality was
still not high enough to tightly constrain the continuum and line
models. Thus, no strong conclusions could be made about the properties
of the inner accretion disk. Firmer results were obtained following a
127-ks \xmm\ observation of 3C~120, the brightest BLRG in the X-ray sky. These data showed a resolved \fe\ line with an
equivalent width of $\sim 50$~eV. The line, however, was clearly
narrow with the innermost radius being $\ga 75$~$r_g$ at an
inclination angle of 10$^{\circ}$.\cite{bfi04,ogle05} This result seemed to
indicate a truncated accretion disk in 3C~120, although the presence
of a highly ionized inner disk could not be ruled out.\cite{bfi04} However,
a very recent observation of 3C 120 by \textit{Suzaku} does show
evidence of a weak (equivalent width $\sim 32$~eV) broad \fe\ line
with an inner radius of $\sim 10$~$r_g$, indicating the presence of an
untruncated optically thick disk.\cite{suz07} Interestingly, the equivalent width
of this line is smaller than the upper limit of any broad component in
a recent simultaneous \textit{Chandra}-\textit{RXTE} observation of
3C~382 that was unable to detect a broad \fe\ line.\cite{glio07} This
result indicates that very long observations may be
necessary in order to detect the broad components of \fe\ lines in
radio-loud AGN. 

Aside from the recent \textit{Suzaku} observation of 3C~120, there
have been two other indications of relativistically broadened \fe\
lines in BLRGs.
\begin{figure}[t!]
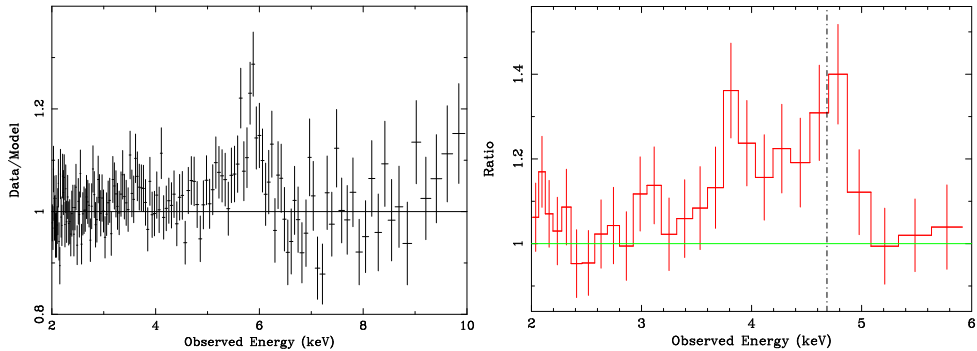

\centerline{\psfig{file=fig2a.eps,width=1.8in,angle=-90}
\psfig{file=fig2b.eps, width=1.8in,angle=-90}}
\vspace*{8pt}
\caption{(Left) The line profile obtained from a \xmm\ observation of
  4C+74.26. The core of the
  line is well fit by narrow \fe\ line, with a weak broad component
  extending to lower energies. The relativistic line must originate
  with 10~$r_g$ and the inner radius is within 6~$r_g$. Plot taken
  from Ref.~\protect\refcite{bf05}. (Right) As in the other panel, but now showing the line
  profile from PG~1425+267. The vertical line indicates the rest-frame
  energy of the neutral \fe\ line at 6.4~\kev. After fitting the
  narrow line core, the remaining broad emission is consistent with a
  relativistic line from the inner disk. Figure taken from
  Ref.~\protect\refcite{mf06}. \protect\label{lines}}
\end{figure}
Figure~\ref{lines} shows the line profiles determined from \xmm\
observations of two broad-line quasars, 4C+74.26 and PG
1425+267. After taking into account the narrow line core at 6.4~\kev, both objects show evidence
for a remaining relativistically broadened component.\cite{bf05,mf06} In the
case of PG~1425+267, the data were not able to precisely determine the
emission radii, but the line is consistent with emission down to
6~$r_g$, the
innermost stable circular orbit (ISCO) of a Schwarzschild (i.e.,
non-spinning) black hole, although the authors were unable to rule out emission
from within this radius.\cite{mf06} In contrast, the inner radius of the broad line in 4C+74.26
could be fit with a value close to the ISCO for a maximally spinning
Kerr black hole.\cite{bf05} In fact, the outer radius was also found to lie
within 10~$r_g$. The reflection continuum was consistent with arising
from a moderately ionized medium. The broadness of these lines and the
complexity of the underlying continua argues for a consistent campaign
of long observations to confirm the presence and extent of these detections.

An alternative, yet highly model dependent, method of probing
accretion geometry in radio-loud AGN is through comparison of spectral
energy distributions with
accretion disk models. When the accretion rate
of an AGN
is less than $\sim 1$\% of its Eddington limit the flow geometry can
transition to an optically-thin, radiatively inefficient flow that
is not expected to provide strong reflection features.\cite{rees82} With the advent
of simple methods to estimate the mass of the central black hole in
AGN, it is now possible to determine the Eddington ratio of large
samples of AGN, modulo
uncertainties in the black hole mass and bolometric
correction. X-ray studies of low-luminosity radio-loud AGN with very
low Eddington ratios suffer from difficulty in isolating the nuclear
continuum from other sources in the galaxy that may be equally
bright. High-resolution imaging by \chandra\ mitigates this problem
and all observations of such low-luminosity radio galaxies have found
no evidence for reflection from an inner accretion
disk.\cite{pan07,bcg06} This supports the hypothesis that these weakly
accreting objects host radiatively inefficient accretion flows
(although see Ref.~\refcite{maoz07}). In contrast, the three AGN with possible relativistic reflection
features in their X-ray spectra (3C~120, 4C+74.26 and PG~1425+267) all
have Eddington ratios $\ga 0.05$, consistent with the idea that these
objects are unlikely to have radiatively inefficient accretion flows
close to the black hole.

In summary, the accretion geometry of radio-loud AGN seems to depend
on the accretion rate. The majority of the radio-loud AGN population
may lie at low accretion rates, where it is expected that there is an
optically-thin, geometrically thick accretion flow close to the black
hole. However, a small fraction of more rapidly accreting AGN also are
radio-loud and indications from relativistically broadened \fe\ lines
show that these sources have `normal' geometrically thin, optically thick disks down to very small
radii. The X-ray spectral observations are very difficult as the sources are
typically faint and very long exposure times are required to provide the
necessary data. Such observations have been very difficult to
come by and the three detections of relativistically broadened \fe\
lines in BLRGs are therefore still tentative. It is imperative,
therefore, that a campaign of long (200+ ks) observations with \xmm\
or \textit{Suzaku} be performed on all nearby BLRGs to provide more
definitive results on the accretion geometry in radio-loud AGN.

\section{How do Black Holes Make Jets?}
\label{sect:jets}
Given the relative scarcity of radio-loud AGN, it must be very
difficult for accreting black holes to produce powerful collimated
jets. This fact then implies that the jet production mechanism
must depend on a number of different parameters or conditions,
each of which must be fulfilled in order to launch a highly
relativistic jet. In this section we collect together many of the
observational results of the last few years, and, in combination with
recent theoretical work, discuss those parameters that are likely to
be important for jet formation. The X-ray spectroscopic results
described in the previous section will also be taken at
face value, although we again emphasize the need for long follow-up
observations to confirm the broad \fe\ line detections.

\subsection{Black Hole Spin}
\label{sub:spin}
Black holes may possess an angular momentum $J$ that is
often conveniently written in dimensionless form
$a=J/J_{\mathrm{max}}=cJ/GM^2$. In 1977 Blandford and Znajek showed
that the spin energy of black hole could be extracted given an
electromagnetic connection between the hole and a nearby `load'.\cite{bz77} The spin energy may then be the ultimate source for producing
powerful radio jets. This can be seen by noting that the available
energy that is potentially extractable from a maximally spinning hole
would be enough to account for the luminosities of the most powerful
radio-loud quasars.\cite{bland90} Numerical simulations of accretion onto
black holes seem to confirm this scenario, with the power of the
numerical jet increasing significantly with the value of $a$.\cite{mck05,hk06}  

Thus, a possible scenario for the origin of radio-loud AGN
is that only those galaxies harbor rapidly spinning black holes,
while the radio-quiet AGN host only weakly spinning or Schwarzschild
black holes.\cite{ssl07,wc95} This would imply that the majority of AGN (and
therefore galaxies) would have weakly spinning black holes. The
justification for this conclusion arises from calculations of black
hole growth through mergers. Only massive black holes that have had
a merging event with another massive object produce a rapidly spinning
black hole.\cite{wc95,hb03} Indeed, since mergers can occur with any random impact
parameter, a number of smaller mergers will not necessarily
increase the spin of the resulting black hole.\cite{hb03} Such a model is
consistent with the observational result that radio-loud AGN almost
exclusively reside in elliptical-like galaxies with core-like
brightness profiles, which are most
naturally formed following a major merger. However, black holes will
also accrete a significant amount of gas following merger events, and
this will almost always cause the black hole to spin up.\cite{shap05}
\begin{figure}[t!]
\centerline{\psfig{file=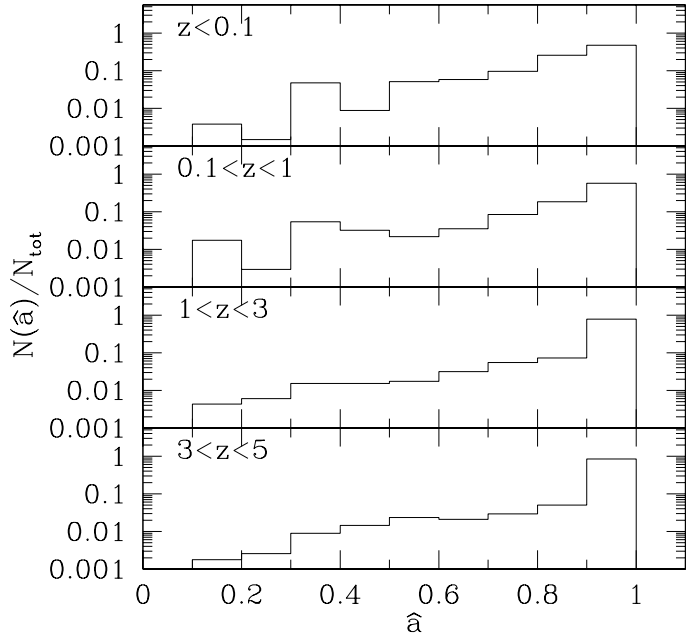,width=3.8in}}
\vspace*{0pt}
\caption{The evolution of the distribution of black hole spins through
  both mergers and accretion. The vast majority of black holes end up
  with high values of $a$. Plot taken from
  Ref.~\protect\refcite{volo05}. \protect\label{spins}}
\end{figure}
Figure~\ref{spins} shows the results of a calculation performed by
Volonteri et al. where they took into account both accretion and
the effects of mergers over cosmological timescales.\cite{volo05} They found that
most black holes are already rapidly spinning by $z \sim 5$. A
possible way to avoid such a rapid increase in $a$ is if a galaxy suffers
very few major mergers and accretes only very small parcels of gas at
random angles.\cite{volo07} In this case, it is possible for counter-aligned
accretion to drive down the angular momentum of the black hole. Such a scenario would lead to radio-quiet AGN residing
primarily in disk galaxies as is observed; however, more work is
required to show that this rather confining scenario is a natural outcome of galaxy evolution.

One other problem with the model of a pure spin dependence on jet
launching arises from the comparison of the local black hole mass
density to the accreted mass density of quasars.\cite{soltan82} All
measurements using the latest quasar luminosity functions find that, in
order to match the accreted mass density to the local quiescent black
hole mass density, the average radiative efficiency of quasars
must be $> 0.1$, higher than the maximum efficiency for a non-spinning
black hole.\cite{elv02,yt02,barg05}. This result implies that the vast majority of black
holes in quasars (which are predominately radio-quiet) have rapidly spinning
black holes.

Observationally, the only direct probe of the spin of a black hole is
through relativistic broadened \fe\ lines. It is therefore interesting
to consider the case of \mcg. This local radio-quiet Seyfert~1 resides in an E-S0
galaxy and has a black hole mass estimate of $\sim 4\times
10^6$~M$_{\odot}$.\cite{mch05} Its X-ray spectrum shows a very broad \fe\ line that requires
emission from within the ISCO of a Schwarzschild black hole.\cite{fab02} In fact,
a very careful spectral analysis performed by Brenneman and Reynolds derived a lower limit to the black hole spin of $a> 0.989$.\cite{br06} This
is to be compared with the tentative detection of the broad \fe\ line in
4C+74.26, that also requires line emission from close to the ISCO of
a rapidly spinning black hole.\cite{bf05} Thus, here are two examples of rapidly
spinning black holes in AGN, one being radio-loud and the other
radio-quiet. Sikora et al. postulate that the example of MCG--6-30-15
(along with radio-quiet quasars) could be explained by the lack of any
appropriate MHD collimation in the inner-region of the accretion
flow.\cite{ssl07} This is similar to arguments that radio-loud
quasars may be equivalent to the flaring states of microquasars and
therefore radio-loud quasars are only a short-lived phenomenon in the
life of a quasar.\cite{nbb05}

Finally, it is worth noting that Galactic black holes typically
produce significant radio emission only in their low/hard
states.\cite{fen01} At higher accretion rates, in their high/soft state, the jet
emission disappears.\cite{fen99,cor00} Spin estimates of Galactic black holes vary
depending on the measurement technique, with broad \fe\ lines
resulting in values of $a \ga 0.9$,\cite{mill04} while comparisons of the accretion
disk temperature and luminosity with spectral models yield $a \la
0.8$.\cite{ddb06} The actual value of $a$ is irrelevant however, as the spin of a
Galactic black hole will not change when it transitions from the low/hard to the high/soft state. Thus, for the Galactic black
holes there must be some other parameter in addition to black hole spin that is required to produce
jetted radio emission.

All these results, both observational and theoretical, seem to suggest
that the spin energy of a black hole is the ultimate energy source for
powerful radio jets, and that therefore a rapidly spinning black hole is a
necessary condition for jet production. However, it is not a
sufficient condition and there must be other parameters that come into
the equation.

\subsection{Accretion Rate}
\label{sub:mdot}
The sensitivity of jet production on the accretion rate seen in
Galactic black holes seems to also be reflected in AGN
data: assuming a constant $R$ dividing line, a larger fraction of AGN
are radio-loud at low accretion rates than at higher ones.\cite{ho02,ghu06,ssl07} Yet, while
jet formation is quenched in Galactic sources once the accretion rate
exceeds $\sim 2$\% of the Eddington limit,\cite{mac03} there are many radio-loud
quasars and BLRGs that are counter-examples in the AGN community. Both 4C+74.26 and \mcg\ have accretion rates greater than this
limit.\cite{bf05,mch05} Therefore, since radio-loud AGN may be found over a wide range
of accretion rates (relative to the Eddington rate), it seems that
this is not a vital parameter in the production of powerful radio
jets, although a greater fraction of the accretion energy may be
directed into radio structures at low accretion rates.

\subsection{Accretion Physics}
\label{sub:disk}
Theoretical studies of jet formation from accreting
black holes have emphasized the importance of the poloidal component
of the magnetic field at the inner edge of the accretion flow in
providing conditions appropriate to jet launching and collimation.\cite{lop99,mei01,rgb06} A simple way to enhance this component of the magnetic field
is to increase the value of $H/r$ of the accretion flow, where $H$ is the
disk scale height and $r$ is the radius along the disk. The
importance of $H/r$ in extracting energy from a spinning black hole
and launching a jet is now being observed in the latest numerical
simulations.\cite{hk06}

Standard radiatively efficient accretion disks are thin with $H/r \ll
1$.\cite{ss73} However, when the accretion rate is very low compared to
Eddington, the flow can transition to a radiatively inefficient phase
that, due to its inability to cool, will have $H/r \sim 1$. Such
accretion flows are expected to exist within low-luminosity AGN and
Galactic black holes in their low/hard states and can then help
explain the presence of radio jets in these objects.\cite{mei01} In this case, all that is
required to produce a powerful radio jet would be a rapidly spinning
black hole to provide the energy, and a favorable magnetic field
geometry to extract this energy and provide collimation. 

We are left then with the problem of radio-loud quasars and BLRGs that
will have rapidly spinning black holes, but will be accreting at too
high a rate to support a radiatively inefficient accretion flow. At these large
accretion rates, the disk may develop a radiation pressure dominated
region close to the black hole that will increase $H/r$.\cite{ss73}
However, both 4C+74.26 and \mcg\ will have such radiation pressure
dominated regions in their accretion disks, but only 4C+74.26 is
successful in producing a large-scale jet.

\subsection{Black Hole Mass}
\label{sub:mass}
Although the value of the black hole mass is not important for jet
formation by Galactic black holes, it does seem relevant for AGN (see
right-hand panel of Fig.~\ref{sikora}). Specifically, the black hole
masses at the center of radio-loud AGN are almost always $>
10^8$~M$_{\odot}$. Radio-quiet AGN can of course also harbor such
massive black holes, but radio-loudness is practically confined to
this range of masses. This holds true for our two test cases: the
radio-quiet \mcg\ has a black hole mass $\sim 4\times
10^6$~M$_{\odot}$, while the black hole in the BLRG 4C+74.26 is
estimated to be 1000$\times$ larger.\cite{wu02}

Clearly, radio-loud AGN do prefer to reside in galaxies that have
undergone a major merger and will therefore harbor the most massive
black holes.\cite{cp06,cp07} These black holes will most likely have a
very large value of $a$, but as discussed in \S~\ref{sub:spin}, this
is not the key parameter in jet formation.  Therefore, the question
still remains: why does the energy flow at the center of the accretion
disk care about the black hole mass? Studies of X-ray variability from
both Galactic black holes and AGN show that, as best as can be
determined, accretion physics is scale free.\cite{mch06} As the black
hole mass determines only the size scale of the accretion system, it
should not be important to the underlying physics. Moreover, what
determines this almost magical dividing line of $\sim
10^8$~M$_{\odot}$? Is it a selection effect, or evidence of the key to
jet formation?

According to simple analytical accretion disk models, $H/r$ is
virtually independent of the central black hole mass,\cite{ss73} but
radiation pressure dominated regions are well known to be unstable in
a variety of different ways.\cite{le74,ss76,bkb77,gam98,bs01}
Simulations of jet formation from radiation pressure dominated disks
have yet to be performed. Perhaps it is possible that the rapid
instabilities in these type of flows discourage jet formation in lower
mass black holes. Alternatively, theoretical models of jet
formation have argued that the total jet power produced is dependent on the
black hole mass.\cite{mei01} Such a dependence has been claimed in
recent studies of certain AGN samples.\cite{ljg06} In this scenario, the total power available to
the jet is much larger in 4C+74.26 than in \mcg, allowing the jet from
4C+74.26 to break out of the host galaxy and become visible in the
low-density intra-galactic medium. More work is clearly
needed to elucidate the connection between black hole mass and jet
formation.

\section{Conclusions}
\label{sect:concl}
Uncovering the mechanisms underlying the production of powerful and collimated radio jets is a very
interesting problem. Jets appear in only a small fraction of AGN, yet
they are a nearly universal phenomenon, as they appear in all
relativistic accreting systems down to neutron stars. The physics
behind this process must at the same time be highly scale free and
adaptable, yet simultaneously be complex enough in order for jets from
AGN to be relatively rare. While a complete understanding of the
processes involved is still lacking, enormous observational progress
over all wavelengths has allowed a narrowing in on the key
parameters.

X-ray spectroscopic measurements of the \fe\ line and reflection
features probes the accretion disk geometry in AGN. There now exists
tentative detections of relativistically broadened \fe\ lines
from three BLRGs: 3C~120, 4C+74.26 and PG~1425+267. These three
examples indicate that these radio-loud AGN have optically-thick,
radiative efficient accretion disks down to small distances from the
black hole, similar to the radio-quiet AGN. It is of paramount
importance that further deep X-ray observations be undertaken to
confirm these detections and expand the sample of known \fe\
lines. In contrast, most radio-loud AGN have such low luminosities
that they will likely harbor geometrically thick, radiatively inefficient accretion
flows. Additional observations and modeling are also needed to confirm
this conclusion.

Combining the multiwavelength data with the results of numerical
studies, we conclude that there are three necessary conditions that
must all be satisfied for the production of a powerful radio jet: a
rapidly spinning black hole (provides the energy source), an inner accretion flow
with a large $H/r$ (provides the enhanced poloidal magnetic field and collimating
mechanism) and a favorable magnetic field geometry. This last
condition is still amorphous, but seems to be required in order to
explain the significant fraction of low-luminosity radio-quiet
AGN. Depending on its individual history, a rapidly spinning black
hole may exist in any galaxy, but seems most likely to reside in a
massive galaxy that has undergone at least one major merger. An
accretion flow with a large $H/r$ is naturally predicted by models of
radiative inefficient accretion flows, but may also be provided by the
radiation pressure dominated region of a standard Shakura-Sunyaev
disk. The black hole mass seems to play an important role in the
formation of jets in radio-loud AGN, perhaps through increasing the total power output
in the jet. 

The evident complexity of this process means that several
reasons are possible to explain why radio-quiet AGN lack a powerful
jet. In the case of radio-quiet Seyfert galaxies that reside in disk galaxies,
the reason may be a lack of a rapidly spinning black hole. In the case
of \mcg, the reason may be that the relatively small black hole mass
does not generate enough jet power. The lack of a suitable magnetic
configuration may explain the radio-quiet quasars. These conditions
can be isolated only by a careful comparison of radio-loud and
radio-quiet AGN at a number of different accretion rates. Finally,
X-ray spectroscopy is currently the only direct observational
technique available to measure black
hole spins. Thus, long X-ray observations are crucial for
understanding the accretion geometry of radio-loud active galaxies.

\section*{Acknowledgments}

DRB is supported by the University of Arizona Theoretical Astrophysics
Program Prize Postdoctoral Fellowship. The author thanks L.\ Ho for
comments on a draft of the manuscript.

%

\end{document}